\documentclass[useAMS,galley]{mn2e}
\usepackage{graphicx}
\usepackage{times,mathptm,lscape}
\def\micron{\hbox{$\mu$m}}
\newcommand{\ltsima}{$\; \buildrel < \over \sim \;$}
\newcommand{\simlt}{\lower.5ex\hbox{\ltsima}}
\newcommand{\gtsima}{$\; \buildrel > \over \sim \;$}
\newcommand{\simgt}{\lower.5ex\hbox{\gtsima}}

\newcommand{\lum}{\rm erg s$^{-1}$}
\newcommand{\lfir}{\rm $L_{\rm FIR}$}
\newcommand{\loh}{\rm $L_{\rm OH}$}
\newcommand{\pn}{\par\noindent}

\def\lesssim{\mathrel{\hbox{\rlap{\hbox{\lower4pt\hbox{$\sim$}}}\hbox{$<$}}}}
\def\gtrsim{\mathrel{\hbox{\rlap{\hbox{\lower4pt\hbox{$\sim$}}}\hbox{$>$}}}}

\def\arcsec{\hbox{$^{\prime\prime}$}}

\def\ab1450{$AB_{1450(1+z)}$}

\def\xray{\hbox{X-ray}}

\def\oiii{\hbox{[O\ {\sc iii}}]}
\def\lsun{\hbox{$L_\odot$}}

\def\h2o{H$_{2}$O}
\def\chandra{{\it Chandra\/}}

\def\heao1{{\it HEAO-1\/}}
\def\hst{{\it {\it HST}\/}}
\def\iras{{\it IRAS\/}}

\def\rosat{{\it ROSAT\/}}

\def\xmm{{XMM-{\it Newton\/}}}

\def\aj{AJ}
\def\araa{ARA\&A}
\def\apj{ApJ}
\def\apjl{ApJ}
\def\apjs{ApJS}

\def\apss{Ap\&SS}
\def\aap{A\&A}

\def\mnras{MNRAS}
\def\nat{Nature}

\def\pasa{PASA} 
 
\def\procspie{Proc.~SPIE}

\title[THE X-RAY PROPERTIES OF OH MEGAMASER SOURCES]
{On the X-ray Properties of OH Megamaser Sources: Chandra Snapshot 
Observations}
\author[C. Vignali et al.]
{
Cristian Vignali,$^{1,2}$\thanks{E-mail: cristian.vignali@bo.astro.it (CV); 
niel@astro.psu.edu (WNB); andrea.comastri@bo.astro.it (AC); 
darling@ociw.edu (JD)}
William N. Brandt,$^{3}$\footnotemark[1] 
Andrea Comastri$^{2}$\footnotemark[1] and 
Jeremy Darling$^{4}$\footnotemark[1] \\ \\ 
$^{1}$ Dipartimento di Astronomia, Universit\`a degli Studi di Bologna, 
Via Ranzani 1, 40127 Bologna, Italy \\
$^{2}$ INAF -- Osservatorio Astronomico di Bologna, Via Ranzani 1, 
40127 Bologna, Italy \\
$^{3}$ Department of Astronomy and Astrophysics, The Pennsylvania 
State University, 525 Davey Laboratory, University Park, PA 16802, USA \\
$^{4}$ Carnegie Observatories, 813 Santa Barbara Street, Pasadena, CA 91101, 
USA \\
}
\begin{document}

\date{Accepted 2005 August 22. Received 2005 July 20}

\pagerange{\pageref{firstpage}--\pageref{lastpage}} \pubyear{2005}

\maketitle

\label{firstpage}

\begin{abstract}
We present \chandra\ snapshot observations for a sample of 7 sources selected 
from the Arecibo OH megamaser (OHM) survey at $z\approx0.13-0.22$ and 
with far-infrared luminosities in excess of 10$^{11}$~\lsun. 
In contrast with the known \h2o\ megamasers, which are mostly associated with 
powerful Active Galactic Nuclei (AGN), the situation is far less clear for 
OHMs, which have been poorly studied in the \xray\ band thus far. 
All of the observed sources are \xray\ weak, with only one OHM, 
IRAS~FSC~03521$+$0028 \hbox{($z=0.15$)}, being detected by \chandra\ 
(with 5 counts). 
The results from this pilot program indicate that the \xray\ emission, 
with luminosities of less than $\approx10^{42}$~\lum, is consistent with 
that from star formation (as also suggested in some cases by the optical 
spectra) and low-luminosity AGN emission. 
If an AGN is present, its contribution to the broad-band emission of 
OHM galaxies is likely modest. 
Under reasonable assumptions about the intrinsic \xray\ spectral shape, 
the observed count distribution from stacking 
analysis suggests absorption of $\approx10^{22}$~cm$^{-2}$. 
\end{abstract}


\section{Introduction}
Extragalactic hydroxyl (OH) megamaser (OHM) emission has been studied 
since the early eighties when it was discovered in Arp~220 
(Baan, Wood \& Haschick 1982). 
To be produced, OHM activity requires 
(1) high molecular density ($n_{H_2}=10^{4-7}$~cm$^{-3}$; Baan 1991), 
(2) a ``pump'' to invert the hyperfine population of the OH ground state, and 
(3) a source of 18~cm continuum emission to stimulate maser emission 
(Burdyuzha \& Komberg 1990), whose main lines are at 1667~MHz and 1665~MHz and 
have luminosities \loh=10$^{1-4}$~\lsun. 
The galaxy merger environment is able to supply all of these requirements: 
the merger/interaction concentrates molecular gas in the merger nuclei, 
creates strong far-infrared (FIR; \hbox{8--1000} \micron) dust emission from 
reprocessed starburst and active galactic nucleus (AGN) activity, 
and produces radio continuum emission. 
The FIR radiation field and/or collisional shocks in the molecular gas can 
invert the OH population and thus allow maser emission. 

Not surprisingly, all known OHMs have been observed in luminous infrared 
galaxies (LIRGs; \hbox{\lfir$>10^{11}$~\lsun}), often favouring the most 
FIR luminous (e.g., Baan 1991; Baan, Salzer \& LeWinter 1998; 
Darling \& Giovanelli 2002a, hereafter DG02a; Lo 2005), 
the ultraluminous infrared galaxies (ULIRGs; \hbox{\lfir$>10^{12}$~\lsun}). 
The high FIR luminosities of ULIRGs are commonly thought to arise from dust 
absorption and FIR re-emission of an intense but obscured starburst and/or 
AGN radiation field. 
While it was known early on that ULIRGs nearly always show evidence for 
interactions (i.e., collisions/mergers; e.g., Sanders \& Mirabel 1996; 
Clements et al. 1996; Borne et al. 2000; Farrah et al. 2001, 2003), 
many investigators have debated 
the luminosity dependence of the fraction of interacting systems among the 
ULIRG population and the possibility that these sources may represent the 
transition stage between galaxy mergers and quasars 
(e.g., Farrah et al. 2001; Tacconi et al. 2002; 
see also Yun et al. 2004). 

OHM studies provide powerful diagnostics of the physical conditions in the 
innermost regions of luminous FIR galaxies. There is evidence 
that the OH emission is often produced by an ensemble of many masing regions 
in the nuclear regions of (U)LIRGs on scales of a few hundred 
parsecs or less (Diamond et al. 1999). 
VLBI observations of a few nearby OHMs have demonstrated that 
OH maser emission can arise in circumnuclear discs or tori 
such as for III Zw 35 (Pihlstr\"{o}m et al. 2001) or Mrk 231 
(Kl\"{o}ckner, Baan \& Garrett 2003), or,
as for Arp~220 (Rovilos et al. 2003),
the emission can show a complicated, irregular morphology and velocity 
structure.  
Blueshifted/redshifted line components in several 
OHMs (Baan, Haschick \& Henkel 1989; Pihlstr\"{o}m et al. 2005) 
have been interpreted as due to bulk outflows/inflows of molecular gas; 
this is 
expected given the tidal streaming and starburst-driven winds associated
with major mergers. 
It is noteworthy that most VLBI studies of OHMs show unresolved maser 
spots of surprising velocity width (tens of km s$^{-1}$), often attributed to 
turbulence and suggestive that compact OHM emission may not be pumped by 
the FIR radiation field alone (e.g. Lonsdale et al. 1998).  
A notable exception is the infrared quasar Mrk 231, which lacks 
a compact maser component contrary to expectation
(Lonsdale et al. 2003; Kl\"{o}ckner, Baan \& Garrett 2003). 
The relationship of OHMs to starbursts versus AGN thus remains murky: all 
manner of masing is seen coupled with all manner of optical classifications.
To obtain deeper insight into the OHM phenomenon, we require 
a minimal-obscuration means to quantify the role of AGN versus starburst
in masing regions.  The radio-FIR relation hints at some difference
between masing and non-masing (U)LIRGs (DG02a), but there is a degeneracy 
between starbursts and AGN in this relation. 
\begin{table*}
\centering
\begin{minipage}{130mm}
\caption{\chandra\ Cycles~4 and 5 observations of OH megamaser galaxies: 
observation log.}
\footnotesize
\begin{tabular}{ccccccc}
\hline
Src. Name & & \multicolumn{2}{c}{Optical} & X-ray Obs. & Exp. Time & Reference \\
\cline{3-4} \\
IRAS FSC  & $z$ & $\alpha_{2000}$ & $\delta_{2000}$ & Date & (ks) & (Arecibo Survey) \\
\hline
01562$+$2528 & 0.1658 & 01 59 02.61 & $+$25 42 35.4 & 2003 Nov. 08--09 & 4.00 & (1) \\
02524$+$2046 & 0.1815 & 02 55 17.09 & $+$20 58 56.5 & 2003 Nov. 08     & 4.31 & (1) \\
03521$+$0028 & 0.1522 & 03 54 42.21 & $+$00 37 03.8 & 2002 Dec. 25     & 9.78 & (1) \\
08201$+$2801 & 0.1680 & 08 23 12.62 & $+$27 51 39.6 & 2003 Nov. 20     & 4.41 & (2) \\
08279$+$0956 & 0.2085 & 08 30 39.38 & $+$09 46 36.1 & 2004 Jan. 13     & 4.38 & (2) \\
09531$+$1430 & 0.2151 & 09 55 50.11 & $+$14 16 07.9 & 2003 Dec. 09     & 4.03 & (2) \\
09539$+$0857 & 0.1290 & 09 56 34.30 & $+$08 43 06.1 & 2004 Jan. 06     & 4.81 & (2) \\
\hline
\end{tabular}
\end{minipage}
\begin{minipage}{130mm}
{\sc Notes ---} The optical positions have been derived from the Digital 
Palomar Sky Survey (DPOSS2) $R$-band images using {\sc SExtractor} (Bertin 
\& Arnouts 1996). 
{\sc References ---} (1) Darling \& Giovanelli 2002a (DG02a); 
(2) Darling \& Giovanelli 2001 (DG01). 
\end{minipage}
\end{table*}


In contrast with H$_{2}$O megamasers, which are likely pumped by an 
AGN \xray\ radiation field (e.g., Baan 1997; Braatz, Wilson 
\& Henkel 1997; Henkel et al. 1998, 2005; Townsend et al. 2001; 
Maloney 2002; Braatz et al. 2003, 2004), 
the AGN content of OHMs has not been investigated with a minimal-obscuration 
probe of AGN activity, aside from radio continuum
observations of radio-loud AGN that lie well off of the radio-FIR relation. 
A previous optical study of OH maser sources (most of which were OHMs) 
found comparable fractions of AGN- and starburst-dominated galaxies (Baan et 
al. 1998) and a non-negligible fraction of composite spectra (i.e., showing 
evidence of both AGN and starburst activity) in the OH maser population.  
Because of their penetrating nature, \hbox{X-rays} represent an efficient tool 
to provide constraints on the engine powering OHMs. Furthermore, X-rays 
typically provide maximal contrast between the AGN emission and that of 
the host galaxy and/or starburst component (e.g., Vignati et al. 1999). \\
Hereafter we adopt $H_{0}$=75~km~s$^{-1}$~Mpc$^{-1}$ in a $\Lambda$-cosmology 
($\Omega_{\rm M}$=0.3 and $\Omega_{\Lambda}$=0.7, Spergel et al. 2003).

\section{Sample selection and source properties}
The goal of the present paper is to provide basic constraints on the \xray\ 
emission of OHM galaxies using \chandra\ snapshot observations. 
Past studies have shown evidence for a correlation between the OH 
line width and the \xray\ luminosity in the soft band (Kandalyan 2003) 
using a compilation of OH masers from the literature 
(mostly from Baan et al. 1998) detected by \rosat. 
This was interpreted as suggesting that \xray\ heating of molecular gas 
provides additional collisional excitation to the maser emission, 
although further studies are required to investigate this issue. 
We note, however, that the compilation of OHMs chosen by Kandalyan (2003) 
is heterogeneous, since it was not drawn from a well defined sample. 
For these reasons, we prefer to focus, in this pilot \chandra\ program, 
on the sample of OHMs selected from the Arecibo Observatory OH megamaser 
survey 
which is a flux-limited survey designed to quantify the relationships 
between merging galaxies and the OHMs 
that they produce, with the goal of using OHMs as luminous tracers of mergers 
(and, possibly, of star formation; Townsend et al. 2001) at high redshifts 
(Darling \& Giovanelli 2000, 2001, 2002b, hereafter DG00, DG01, and 
DG02b, respectively). 
This survey has been conducted over about one-quarter of the sky to a 
distance limit of 
roughly 1~Gpc. Candidates for observations with the Arecibo Observatory were 
selected from the PSCz (Saunders et al. 2000), which is a flux-limited 
(\iras\ $f_{60~\micron}>0.6$~Jy) redshift survey of $\approx$~15,000 
\iras\ galaxies over 84 per cent of the sky. 
The Arecibo survey detected 50 new OHMs in the redshift range 
\hbox{0.11--0.27} (DG02a,b), thus doubling the sample of known OHMs and 
increasing the $z>0.1$ sample sevenfold. Thanks to this survey, it is now 
evident that the fraction of OHMs in (U)LIRGs is an increasing function of 
\lfir: about one-third of the ``warmest'' (i.e., with higher dust 
temperatures) ULIRGs are characterized by OH megamasers (DG02a). 

The OHM galaxies proposed for observation with \chandra\ have been chosen 
using the catalog published by DG01 available at the time of \chandra\ 
Cycle~5 proposals (6 of the 7 OHMs described in the following); 
our targets are among the highest luminosity members 
(\loh\ $>10^{3}$~\lsun; $\approx$~30--40 per cent of the OHM population in the 
Arecibo survey) of the maser population (filled circles in Fig.~1). 
%
\begin{figure}
\includegraphics[angle=0,width=85mm]{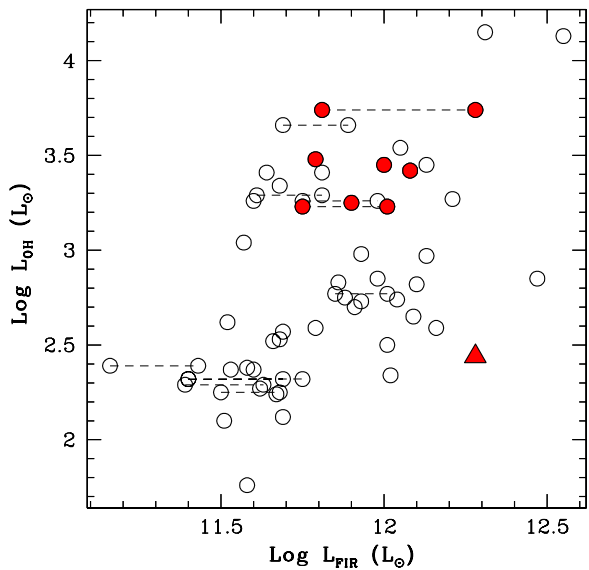}
\caption{
Logarithm of the OH luminosity vs. FIR luminosity (both in units of 
\lsun) for the Arecibo OHM final sample (DG02b). 
The filled symbols indicate the OHMs with \xray\ observations, while the 
filled triangle shows the \xray\ detected OHM. 
The short-dashed lines indicate the permitted range of \lfir\ when 
the flux density at 100~\micron\ is only available as an upper limit 
(see DG02b); in the case of OHMs with \xray\ observations, both circles 
are filled. Note that all of the OHM galaxies have \lfir\ in the LIRG regime, 
with a significant fraction ($\approx$~35 per cent) being ULIRGs.}
\label{fig1}
\end{figure}
They also constitute 
a complete sample, since they were targeted in order of right ascension. 
One additional OHM, characterized by lower OH luminosity 
(filled triangle in Fig.~1), 
has been included using an archival \chandra\ observation. 
It is worth noting that our OHMs lie at higher redshifts than most of the 
OHMs with \xray\ detections in the literature (e.g., Arp~220 at $z=0.018$, 
Mrk~231 at $z=0.042$, etc.). 
On the basis of the merger/interaction scenario for OHMs discussed in $\S$1, 
the present sample, with all of the 7 OHMs being characterized by double 
nuclei\footnote{We note 
that, given the presence of double nuclei, the optical positions reported in 
Table~1 are intermediate between the two nuclei; in these cases, both nuclei 
fall in the same 2\arcsec-radius aperture used for \xray\ manual photometry 
(see $\S$3). 
For IRAS~FSC~01562$+$2528, we report the position of the optically most 
luminous nucleus, given the association of the OH and radio emissions with 
this nucleus.}
and/or tidal tails (Darling et al., in preparation), 
can be considered representative 
of the OH megamaser population overall.
Furthermore, their IR colour ($\log\ f_{100}/f_{60}$) distribution shows no 
significant evidence for any observational bias toward ``warm'' 
(i.e., lower $f_{100}/f_{60}$ values, possibly related to a 
contribution from an active nucleus) or ``cold'' galaxies (see Fig.~2). 
%
\begin{figure}
\includegraphics[angle=0,width=85mm]{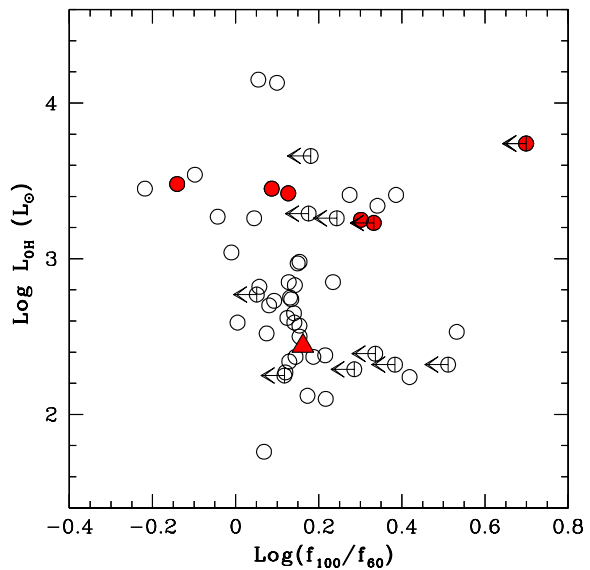}
\caption{
Logarithm of the OH luminosity (in units of \lsun) vs. IR colour 
($\log\ f_{100}/f_{60}$) for the OHM galaxies from DG02b. 
Symbols are the same as in Fig.~1.}
\label{fig2}
\end{figure}

\subsection{Notes on individual sources and optical spectra}
In the following, 
we report on the OH line and optical properties of the 7 OHM 
galaxies with \chandra\ observations. A detailed description of their 
optical spectra is presented in Darling et al. (in preparation). 
%
\begin{description}
\item {\sf 01562$+$2528}: this source shows the presence of two nuclei 
(one much brighter than the other) in the optical (with a separation of 
6.3\arcsec, i.e., 17.9~kpc), suggestive of a multiple merger. 
The Eastern component has 
a starburst classification, while for the Western component there is not 
enough signal for a spectral classification (see Table~2 for references). 
\item {\sf 02524$+$2046}: this object has the most unusual spectrum of the 
Arecibo OHM survey, showing multiple strong narrow components in both OH 
lines. The optical morphology is elliptical-like with a single tidal tail 
(Darling et al., in preparation), while the optical spectrum is typical of 
a starburst galaxy. 
\item {\sf 03521$+$0028}: this source is comprised of a close pair; both 
components, separated by 1.3\arcsec\ (corresponding to 3.4~kpc), 
are visible in the Sloan Digital Sky Survey (SDSS) but not in the DPOSS2. 
This object has been classified as a starburst according to the ISO spectrum 
(Lutz, Veilleux \& Genzel 1999) and a low-ionisation nuclear emission-line 
region galaxy (LINER) according to the optical spectrum 
(Nagar et al. 2003).
\item {\sf 08201$+$2801}: the \hst\ image shows two nuclei (with a separation 
of 1.1\arcsec, corresponding to 3.2~kpc), having a starburst classification, 
and a tidal tail, suggesting an advanced merger (DG01). 
The NVSS flux density reported in DG01 is associated with a source at a 
distance of $\approx$~23\arcsec. Nagar et al. (2003) reported a flux-density 
upper limit of 0.8~mJy at 15~GHz, roughly corresponding to a 
\hbox{1.4--15~GHz} spectral slope of $\simgt$~0.75. 
\item {\sf 08279$+$0956}: the optical image of this source (classified as 
a LINER) shows the presence of 3 tidal tails (Darling et al., in preparation). 
\item {\sf 09531$+$1430}: this galaxy is an interacting system whose 
components are blended in the images from the Automatic Plate Measuring (APM) 
facility; 
the 2MASS position is on the Western component (having a Seyfert~2 spectrum), 
while the FIRST position is on the Eastern component (characterized by a 
starburst spectrum; Darling et al., in preparation). 
\item {\sf 09539$+$0857}: the optical image shows the presence of tidal tails 
and a single nucleus, classified as a LINER (Veilleux, Kim \& Sanders 1999) 
or Seyfert~2 (Darling et al., in preparation). 
The two classifications were performed in an identical manner. 
In both cases, the \oiii/H$\beta$ ratio lies very close to the (somewhat 
arbitrary) dividing line between LINERs and Seyfert~2s, and the measured line 
ratios obtained by both groups agree to within the measurement uncertainties.
%
\begin{table*}
\begin{minipage}{\textwidth}
\caption{Multi-wavelength properties of the Arecibo sample of OHMs observed 
by \chandra.}
\tiny
\begin{tabular}{ccccccccccccccccccc}
\hline
Source & $z$ & $N_{\rm H}$ & $R$ & $B$ & & $f_{12}$ & $f_{25}$ & $f_{60}$ & $f_{100}$ & 
$\log L_{\rm FIR}$ & $\log L_{\rm OH}$ & $f_{\rm 1.4~GHz}$ & $f_{\rm 1.4~GHz}$ & $F_{\rm FB}$ & $\log L_{2-10}$ & Class. & Ref. \\ \\
\cline{4-5} \cline{7-10} \cline{13-14} \\ 
  &  &  & \multicolumn{2}{c}{APM} & & \multicolumn{4}{c}{\iras} & & & FIRST & NVSS &  &  &  &  \\ \\
(1) & (2) & (3) & (4) & (5) &  & (6) & (7) & (8) & (9) & (10) &  (11) & (12) & (13) & (14) & (15) & (16) & 
(17) \\
\hline
01562$+$2528          & 0.1658 & 7.46 & 15.58 &    17.23 & & $<0.126$ & $<0.133$ & 0.809    & 1.623    & 11.90        & 3.25 & \dotfill & 5.9      & 
                      $<11.3$             & $<4.4\times10^{41}$ & {\sc STB}              & (1)    \\
02524$+$2046          & 0.1815 & 11.3 & 18.08 &    20.02 & & $<0.055$ & $<0.079$ & 0.957    & $<4.789$ & 11.81--12.28 & 3.74 & \dotfill & 2.6      & 
                      $<11.4$             & $<5.4\times10^{41}$ & {\sc STB}              & (1)    \\
03521$+$0028$\dagger$ & 0.1522 & 12.6 & 18.44 & $>21.13$ & & $<0.250$ & 0.233    & 2.638    & 3.833    & 12.28        & 2.44 & \dotfill & 6.1      & 
                      3.9$^{+2.7}_{-1.7}$ & $1.3\times10^{41}$  & {\sc STB/L}            & (1,2,3,4) \\
08201$+$2801$\dagger$ & 0.1680 & 3.59 & 17.12 &    18.39 & & $<0.085$ & $<0.162$ & 1.171    & 1.429    & 12.00        & 3.45 & 4.74     & \dotfill & 
                      $<4.4$              & $<1.8\times10^{41}$ & {\sc STB}              & (4,5)   \\
08279$+$0956          & 0.2085 & 3.80 & 18.08 &    20.04 & & $<0.079$ & $<0.192$ & 0.586    & $<1.263$ & 11.75--12.01 & 3.23 & 2.84     & 3.7      &  
                      $<9.5$              & $<6.1\times10^{41}$ & {\sc L}                & (1)     \\
09531$+$1430          & 0.2151 & 3.07 & 17.25 & $>$19.27 & & $<0.140$ & $<0.169$ & $<0.776$ & 1.040    & 12.08        & 3.42 & 3.23     & 2.5      & 
                      $<10.2$             & $<1.7\times10^{42}$ & {\sc S2/L}$\dagger\dagger$ & (1)      \\
09539$+$0857$\dagger$ & 0.1290 & 3.08 & 17.74 &    19.62 & & $<0.154$ & $<0.147$ & 1.438    & 1.044    & 11.79        & 3.48 & 5.32     & 8.50     & 
                      $<8.5$              & $<1.9\times10^{41}$ & {\sc S2/L}             & (1,3)     \\
\hline
\end{tabular}
\scriptsize
Luminosities are computed using H$_{0}$=75~km~s$^{-1}$~Mpc$^{-1}$, 
$\Omega_{\rm M}$=0.3 and $\Omega_{\Lambda}$=0.7. \\
$\dagger$ Also in the SDSS Data Release 3 catalog (DR3; Abazajian et 
al. 2005). \\ 
$\dagger\dagger$ The Eastern nucleus is classified as a starburst, while the 
Western nucleus is classified as a Seyfert~2 (see text for details). \\
{\sf (1)} \iras\ FSC Name; 
{\sf (2)} redshift; 
{\sf (3)} Galactic column density from Dickey \& Lockman (1990), 
in units of $10^{20}$~cm$^{-2}$; 
{\sf (4--5)} $R$- and $B$-band magnitudes from APM; 
{\sf (6--9)} \iras\ flux densities, in units of Jy; 
{\sf (10)} $\log$ of the FIR luminosity (in units of \lsun) computed according to 
\lfir=$3.96\times10^{5}\times\ D_{\rm L}^2\ (2.58\times\ f_{60}+f_{100})$, 
where $D_{\rm L}$ is the luminosity distance in Mpc (see Fullmer \& Lonsdale 1989); 
when $f_{100}$ is an upper limit, the permitted range of \lfir\ is reported (see $\S$2.2 of DG02b); 
{\sf (11)} $\log$ of the OH luminosity, in units of \lsun; 
{\sf (12)} integrated 1.4~GHz flux density from FIRST (Becker, White \& 
Helfand 1995); 
{\sf (13)} 1.4~GHz flux density from NVSS (Condon et al. 1998); 
{\sf (14)} Galactic absorption-corrected flux in the observed 
\hbox{0.5--8~keV} band, in units of $10^{-15}$~erg~cm$^{-2}$~s$^{-1}$; 
{\sf (15)} 2--10~keV rest-frame luminosity corrected for the effects of 
Galactic absorption and computed from the full-band flux (or upper limit) 
assuming $\Gamma=2.0$, in units of erg~s$^{-1}$; 
{\sf (16)} spectroscopic classification: 
STB=starburst; L=LINER; S2=Seyfert~2; 
{\sf (17)} reference for the spectroscopic classification. \\
{\sc List of references ---} 
(1) Darling et al., in preparation; (2) Lutz, Veilleux \& Genzel 1999; 
(3) Veilleux, Kim \& Sanders 1999; (4) Nagar et al. 2003; 
(5) Kim, Veilleux \& Sanders 1998. \\
\end{minipage}
\end{table*}

Nagar et al. (2003) reported a flux density of 1.2~mJy at 15~GHz, 
corresponding to a \hbox{1.4--15~GHz} spectral slope of $\approx$~0.6--0.8. 
\end{description}

\section{X-ray observations and data reduction}
Six out of the 7 OHMs presented here were targeted by \chandra\ during 
Cycle~5. IRAS~FSC~03521$+$0028, found at the aim point of a \chandra\ 
archival observation, was observed in Cycle~4. 
This source was targeted in a \chandra\ program aimed at observing a sample 
of ULIRGs (see Teng et al. 2005); 
we are confident that the inclusion of this object does not bias our 
``original'' sample in terms of AGN content (see $\S$4). 
The observation log is shown in Table~1. 

All of the sources were observed with the Advanced CCD Imaging Spectrometer 
(ACIS; Garmire et al. 2003) with the S3 CCD at the 
aimpoint. Standard data reduction was adopted (see $\S$2 of Vignali 
et al. 2005 for a detailed description) using the \chandra\ Interactive 
Analysis of Observations ({\sc ciao}) Version~3.2 software. 
Source detection was carried out with {\sc wavdetect} (Freeman et al. 2002) 
using a false-positive probability threshold of 10$^{-6}$. 
We searched for \chandra\ sources within $\approx$~0.8\arcsec\ of the optical 
position. 
IRAS~FSC~03521$+$0028 is the only detected source 
(filled triangle in Fig.~1), with 5 counts in 
the observed \hbox{0.5--8~keV} band; we note that this is also the source 
with the highest exposure by \chandra\ (see Table~1). 
Given the adopted threshold and the small number of 
pixels being searched due to the known source positions and the sub-arcsec 
on-axis angular resolution of \chandra, the probability that this detection is 
spurious is extremely low ($\approx8\times10^{-6}$).  
All of the remaining sources (shown as filled circles in Fig.~1) 
were not detected also using higher (i.e., less conservative) 
threshold values (up to 10$^{-4}$). 

Table~3 summarizes the \xray\ photometric results in the soft band 
\hbox{(0.5--2~keV)}, the hard band \hbox{(2--8~keV)}, and the full band 
\hbox{(0.5--8~keV)} using circular apertures centred on the optical source 
position and an extraction radius of 2\arcsec. 
Although two \xray\ photons falling in adjacent pixels 
can be considered in some circumstances a detection (given the low background 
level of \chandra\ observations; see, e.g., Vignali et al. 
2005), we prefer to follow a more conservative approach and report these 
sources (2 in the present sample; see Table~3) as \xray\ undetected.

\begin{table}
\begin{minipage}{85mm}
\caption{X-ray counts and full-band count rates.}
\begin{tabular}{ccccc}
\hline
  & \multicolumn{3}{c}{X-ray Counts} & CR \\
\cline{2-4} \\
Source & (0.5--2~keV) & (2--8~keV) & (0.5--8~keV) & (0.5--8~keV) \\
\hline
01562$+$2528 & $<3.0$              & $<6.4$ & $<6.4$$\star$       & $<1.60$                \\
02524$+$2046 & $<4.8$              & $<4.8$ & $<6.4$              & $<1.49$                \\
03521$+$0028 & 3.0$^{+2.9}_{-1.6}$ & $<6.4$ & 4.9$^{+3.4}_{-2.1}$ & 0.50$^{+0.35}_{-0.22}$ \\
08201$+$2801 & $<3.0$              & $<3.0$ & $<3.0$              & $<0.68$                \\
08279$+$0956 & $<6.4$              & $<3.0$ & $<6.4$$\star$       & $<1.46$                \\
09531$+$1430 & $<3.0$              & $<6.4$ & $<6.4$              & $<1.59$                \\
09539$+$0857 & $<6.4$              & $<3.0$ & $<6.4$              & $<1.33$                \\
\hline
\end{tabular}
$\star$ The two \xray\ counts are contiguous. \\ 
Errors on the \xray\ counts were computed according to Gehrels (1986). 
The upper limits are at the 95\% confidence level and were computed 
according to Kraft, Burrows \& Nousek (1991). 
For the sake of clarity, upper limits of 3.0, 4.8, and 6.4 indicate that 
0, 1, and 2 \xray\ counts, respectively, have been found within an extraction 
region of radius 2\arcsec\ centred on the position of the OHM source 
(considering the background within this source extraction region to be 
negligible). Count rates (CR) are in units of $10^{-3}$ counts~s$^{-1}$.
\end{minipage}
\end{table}

The final catalog of OHMs from the Arecibo survey (DG02b) 
was also cross correlated with the \rosat\ and \xmm\ archives; 
we obtained no further \xray\ detections. 
\xray\ constraints from \rosat\ observations (available for 9 OHMs) are 
generally loose because of the large off-axis angles of the sources and the 
shallow exposures of the pointed PSPC and HRI observations; hence these 
constraints will not be used in this paper. 
One OHM was targeted by \xmm\ (IRAS~FSC~13218$+$0552 at \hbox{$z=0.205$}, with 
net exposures of $\approx$~8.6--9.7~ks in the 
EPIC MOS/pn cameras) but only an upper limit, consistent with the values we 
obtained for our sources (see $\S$4 and Table~2), was derived 
(Bianchi et al. 2005). Recently, it has been suggested that this source, 
characterized by an extremely red infrared continuum (Low et al. 1989) and 
a blueshifted component in the \oiii\ line (e.g., Zheng et al. 2002), might 
harbour a Compton-thick AGN (i.e., with $N_{\rm H}\gtrsim10^{24}$~cm$^{-2}$), 
surrounded by a strong starburst (see $\S$4.4 and Fig.~6 of Bianchi et al. 
2005).  
We also note that IRAS~FSC~13218$+$0552 has the warmest $f_{100}/f_{60}$ 
colour in Fig.~2 (i.e., it is the left-most object in the plot), 
suggestive of an AGN contribution at IR wavelengths.

\section{What are X-rays telling us about the nature of the OH megamasers?}
The broad-band properties of the OHM sample targeted by \chandra\ 
are reported in Table~2. 
Both the photon index and \xray\ luminosity can provide 
constraints on the nature of the high-energy emission from OHM galaxies. 
However, such constraints, with the current data, are basic due to the 
low counting statistics. 
All of the \xray\ sources have rest-frame 2--10~keV luminosities at most of 
$\approx$~10$^{42}$~\lum\ (see Table~2), which is consistent with both star 
formation (e.g., Zezas, Alonso-Herrero \& Ward 2001; Alexander et al. 2002) 
and low-luminosity AGN (LLAGN) activity (e.g., Terashima et al. 2002; 
Terashima \& Wilson 2003). 
The possibility that the observed \xray\ emission is largely due 
to star formation finds support from Fig.~3, where the FIR luminosity 
is plotted vs. the 0.5--8~keV luminosity. 
Our sources are located in the left-most 
part of the diagram, where either starbursts (e.g., Arp~220) or 
starburst-dominated (i.e., with little apparent contribution 
from the AGN to the bolometric emission; 
see Alexander et al. 2005a) sub-mm sources typically lie. 
\begin{figure}
\includegraphics[angle=0,width=85mm]{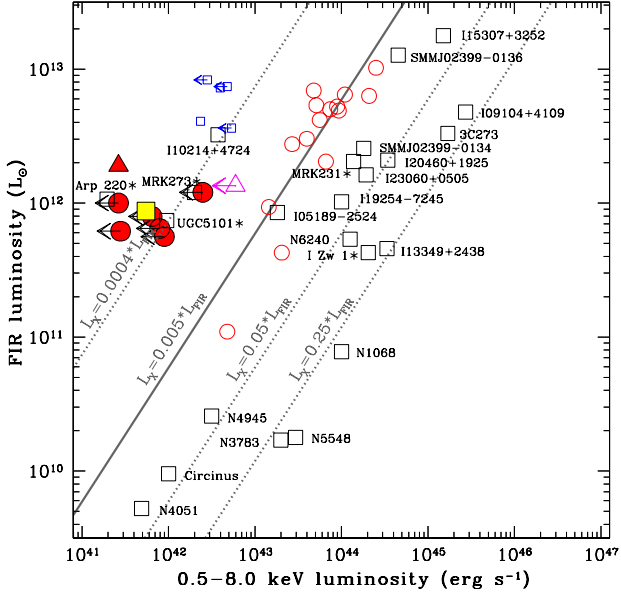}
\caption{
Rest-frame FIR luminosity vs. \xray\ luminosity for the OHM galaxies 
presented in this paper (filled circles: \xray\ upper limits; 
filled triangle: \xray\ detected OHM; filled square: stacked \xray\ counts 
from the whole sample of 7 OHMs observed by \chandra, 
see $\S$4.1). 
The open triangle indicates the OHM from the Arecibo sample which was 
observed but not detected by \xmm\ (see Bianchi et al. 2005 and $\S$3). 
For comparison, some other sources are reported (after correcting, 
when possible, for intrinsic absorption in the \xray\ band): 
the sub-mm galaxies in the \chandra\ Deep Field-North (Alexander et al. 2003, 
2005a,b) likely containing an AGN (open circles) and 
a ``pure'' starburst (small open squares); 
well known AGN-dominated (mostly in the right-hand side of the plot)
and star-formation dominated sources (in the left-most region of the figure) 
from the literature plus two distant well-studied sub-mm galaxies 
(SMM~02399$-$0134 and SMM~02399$-$0136), all shown as large open squares with 
labels. 
The slanted lines show ratios of constant \xray\ to FIR luminosity; this ratio 
is typically $\approx$~3--30 per cent for AGN-dominated sources. Adapted from 
Alexander et al. (2003); see Alexander et al. (2005a) for details. 
Note that some of the sources from the literature sample are low-redshift 
OHM galaxies (marked by a ``$\ast$'' after the source name).}
\label{fig3}
\end{figure}
However, some \xray\ obscured AGN whose broad-band emission is dominated 
by star formation are also present in this region; a fraction of these 
sources are lower redshift OHM sources (e.g., Mrk~273 and UGC~5101); 
the possibility that our sources could contain absorbed AGN will be discussed 
in detail using \xray\ stacking analysis (see $\S$4.1). 

Optical/near-infrared (NIR) spectroscopy (see $\S2.1$) 
indicates that 50 per cent of our sources show the 
typical signatures of starburst galaxies, 25 per cent have a LINER 
classification, and 25 per cent a Seyfert~2 classification 
(according to the classification reported in Darling et al., in preparation; 
see also column (16) in Table~2); 
IRAS~FSC~09531$+$1430 is counted twice because both nuclei were 
spectroscopically identified (see Darling et al., in preparation). 
We note, however, that the optical classification of these sources is 
difficult: when more than one nucleus is present, unless the projected 
physical separation of the nuclei is larger than the resolution of the 
spectrograph, 
optical spectra will be a blend of nuclei, dominated by the optically 
dominant (i.e., the least obscured and/or most optically luminous in emission 
lines) nucleus. 
However, OHMs seem to favour the dustiest environments and may select the 
optically ``subordinate'' nucleus.  Therefore, there may be little 
correlation between OHM properties and optical spectral type. 
Furthermore, we note that optical and NIR studies are sometimes 
not adequate to unveil the engine of the 
\xray\ emission. An emblematic, though perhaps extreme, case of 
different classification at different wavelengths is provided by the 
ULIRG NGC~6240, which is classified as a starburst in the NIR 
(Genzel et al. 1998), a LINER in the optical (Veilleux et al. 1995) and 
a Compton-thick 
AGN in the hard \xray\ band (Vignati et al. 1999; also see 
Komossa et al. 2003). Some further cases similar to that of NGC~6240 
are presented by Maiolino et al. (2003). 

Finally, it is interesting to note that the \xray\ luminosity 
and the \xray-to-FIR luminosity ratio of the only 
\xray\ detected OHM in the present sample do not appear to be different from 
those of the other sources; given the number of observed counts (see Table~3), 
the detection of this source has been possible because of the exposure time 
which is double that of the other OHMs.

\subsection{X-ray stacking analysis}
We can place tighter constraints on the average \xray\ properties of the OHMs 
under investigation by ``stacking'' their counts. Our sources 
have 8$^{+4.0}_{-2.8}$ counts in the \hbox{0.5--2~keV} band and 
7$^{+3.8}_{-2.6}$ counts in the \hbox{2--8~keV} band. 
Monte Carlo simulations in regions close to our targets 
(see Vignali et al. 2005 for details about the 
adopted procedure) indicate that the background in all of the 7 fields 
does not contribute more than 0.14 and 0.31 counts in the soft and hard bands, 
respectively; 
therefore we are confident we obtain secure detections in 
both bands by stacking the counts from the 7 OHMs.\footnote{The 
Poisson probability of obtaining 8 (7) counts or more when 0.14 (0.31) 
counts are expected is 
\hbox{$\approx3.2\times10^{-12}$} (\hbox{$\approx4.2\times10^{-8}$}).} 
The ratio of the observed hard vs. soft-band source counts 
(0.85$^{+0.65}_{-0.43}$) suggests that our sources have, on average, 
a photon index flatter ($\approx0.7^{+0.6}_{-0.5}$) 
than $\Gamma$=1.8--2.1, which is typically used for AGN and 
starburst galaxies at hard \xray\ energies (e.g., Piconcelli et al. 2005; 
Ptak et al. 1999).\footnote{We note, however, that a large population 
of neutron star high-mass \xray\ binaries, whose presence is expected in cases 
of galaxies undergoing intense star formation, can produce a flat 
($\Gamma\approx$~0.5--1) photon index; see Colbert et al. (2004).} 
If the hardness of our sources is due to the presence of intrinsic absorption, 
the implied average column density is 
$\approx$~1.5$\times10^{22}$~cm$^{-2}$ (ranging between 
$\approx$~5.8$\times10^{21}$~cm$^{-2}$ and 
$\approx$~2.6$\times10^{22}$~cm$^{-2}$) at the average redshift of $z=0.17$. 
The corresponding absorption-corrected rest-frame \hbox{2--10~keV} luminosity 
would be in the range $\approx$~1--3$\times10^{41}$~\lum. 
Intrinsically flatter \xray\ slopes would produce lower column densities 
(by $\approx$~20 per cent, if $\Gamma=1.6$ instead of 2.0 is adopted), 
while the presence of an additional soft \xray\ component 
(e.g., thermal emission from the starburst) would produce the opposite effect. 

Figure~3 shows that, if an active nucleus is present in the 
OHMs under investigation, its contribution to their broad-band emission 
is probably modest, unless the AGN is Compton-thick. 
In the Compton-thick case, the \xray\ emission in the \chandra\ band 
would be the small (presumably a few per cent; see, e.g., Comastri 2004; 
Guainazzi, Matt \& Perola 2005) 
scattered/reflected fraction of the nuclear \xray\ emission, 
which would be $\simlt$~a~few~$\times10^{43}$~\lum\ (i.e., at most in the 
Seyfert regime). The reprocessing of the \xray\ and ultraviolet 
emissions at longer wavelengths would still be insufficient to produce 
the observed FIR luminosity without requiring a dominant 
contribution from star formation. 
We note that some sources from the literature sample are characterized by 
Compton-thick absorption (e.g., Mrk~231, Gallagher et al. 2002, 2005,  
Braito et al. 2004; UGC~5101, Maiolino et al. 2003). 

Although the presence of obscuration is uncertain and needs to be 
confirmed by deeper observations with \chandra\ for an extended sample 
of OHMs and, possibly, with \xmm\ for the most ``promising'' sources, 
the column density 
allowed by the current data is consistent with the \xray\ absorption 
observed in many \h2o\ megamasers. These sources, however, 
are mostly associated with \xray\ bright ($\approx$~10$^{42-43}$~\lum) 
and obscured ($N_{\rm H}\gtrsim10^{22}$~cm$^{-2}$) 
Seyfert~2 galaxies (e.g., Maloney 2002; Braatz et al. 2003), 
where the amplifying water vapor molecules are most likely pumped by 
collisional processes resulting from \xray\ heating in parsec-scale molecular 
tori with densities of $\approx$~10$^{7-8}$~cm$^{-3}$ 
(e.g., Neufeld, Maloney \& Conger 1994). 
Column densities of $\approx$~10$^{22}$~cm$^{-2}$ are not uncommon in local 
(U)LIRGs (e.g., Risaliti et al. 2000) and LLAGN (e.g., Terashima et al. 2002; 
Terashima \& Wilson 2003). 
Given the presence of double nuclei and/or indications of past mergers in the 
OHMs of our sample, it is possible that the starburst itself 
can produce the obscuration, as suggested by some authors 
(e.g., Fabian et al. 1998; Ohsuga \& Unemura 2001; Wada \& Norman 2002).

\subsection{X-ray observations of ``bona-fide'' OHMs from the literature}
We obtained further constraints on the \xray\ emission from OHM galaxies 
using the sample of 54 ``bona-fide'' OH masers 
[both kilo- ({\loh$\leq10^{1}$~\lsun}) and mega-] from the literature 
(mostly from Baan et al. 1998), listed in Tables~7 and 8 of DG02a. 
We caution the reader against over-interpretation of the results we present 
in this section: this sample is not well defined, lacks either published 
OH spectra or measurements of the 1667~MHz OH line in about 50 per cent of 
the cases, and includes some suspect OH detections, as pointed out by DG02a. 

Twenty-four OH masers have accessible pointed \xray\ observations in the 
\rosat, \chandra, and \xmm\ archives (for one additional source, the \xmm\ 
data are not public yet); all of these galaxies are OHMs. 
Seventeen sources are \xray\ detected; the high 
detection rate ($\approx$~70 per~cent) is due to the fact that most of these 
sources were targeted by the sensitive \xray\ instruments on-board \chandra\ 
and \xmm, while the \xray\ non-detections are mostly due to \rosat\ 
observations where the sources are located at large off-axis angles. 
For most of the sources with sensitive observations, \xray\ information is 
available from published work 
(Gallagher et al. 2002, 2005; Xia et al. 2002; Maiolino et al. 2003; 
Ptak et al. 2003; Franceschini et al. 2003, hereafter F03; 
Satyapal, Sambruna \& Dudik 2004; 
Ballo et al. 2004; Braito et al. 2004; Imanishi \& Terashima 2004; 
Iwasawa et al. 2005); for a couple of sources, \chandra\ data have 
been retrieved from the archive and analysed using standard procedures. 

We find that AGN emission (with intrinsic \xray\ luminosities above 
$\approx~10^{42}$~erg~s$^{-1}$) is present in eight OHMs. Four of these OHMs 
are likely characterized by Compton-thick absorption 
(UGC~05101, Maiolino et al. 2003; NGC~3690 in the galaxy pair Arp~299 system, 
Della Ceca et al. 2002, Ballo et al. 2004; NGC~4418, Maiolino et al. 2003; 
Mrk~231, Gallagher et al. 2002, 2005; Braito et al. 2004). 
However, Compton-thick absorption cannot be ruled out in some low 
signal-to-noise ratio sources and in Arp~220 (Iwasawa et al. 2001, 2005). 

To provide a more appropriate comparison with our \chandra\ targets, 
from the subsample of literature OHM with \xray\ coverage, 
we selected those within the same redshift interval, the same 
{\loh} range, or the same {\lfir} range 
as our seven OHMs observed by \chandra. 
The paucity of objects in the same redshift range (two, with only one 
\xray\ detection) does not allow for any reasonable comparison with our 
sample; however, it highlights the relevance of the present work, 
where the redshift range of OHMs with \xray\ constraints has been 
extended significantly. 
If we select the literature OHMs using the same interval of OH or FIR 
luminosities as our targets, we find a larger number of objects (12 and nine, 
respectively). AGN emission is clearly present in three objects 
(one Compton-thick, Mrk~231, and the other two Compton-thin, Mrk~273 and 
PKS~B1345$+$145), corresponding to $\approx$~25--33 per~cent of the chosen 
subsamples. LLAGN could still be present in some of the remaining sources, 
although the \xray\ emission from these objects can be well explained by a 
thermal gaseous component from a starburst plus a hard, moderately absorbed 
power-law component due to the unresolved binary population 
(e.g., F03; Teng et al. 2005). 

As a further test, we compared the \xray\ spectral results for our seven OHMs 
(see $\S$4.1) with those obtained by F03 with \xmm\ for their 
five ``bona-fide'' lower redshift OHMs without a dominant AGN component 
(i.e., excluding Mrk~231). 
We found that our stacked spectrum is generally harder than the \xray\ 
spectra of the F03 OHMs. This result can be ascribed to the different 
spatial resolutions of the \chandra\ ACIS and \xmm\ EPIC cameras and 
different source-extraction regions. 
To test this possibility, we used a sample of three starburst galaxies 
observed by \chandra\ (Ptak et al. 2003) and \xmm\ (F03). The soft
thermal component appears spatially more extended than the hard component, 
i.e., it likely provides a stronger contribution to the broad-band \xray\ 
spectrum if large source-extraction regions are chosen (as in the case of 
\xmm\ spectra, extracted from circular regions of radius 20\arcsec). 
Motivated by these considerations, we rescaled the source-extraction regions 
of our OHMs to match those adopted by F03 (taking into account the different 
redshift range of the two samples), obtaining a somewhat softer \xray\ 
spectrum ($\Gamma\approx1.6\pm{0.3}$), consistent with F03 results. 

In the light of the results obtained from the study of the literature OHMs 
and keeping in mind all the caveats described above, starburst emission with 
moderate absorption probably within the binary population 
is likely to explain the average \xray\ properties 
of our seven OHMs ($\S$4.1).

\section{Summary}
We have analysed \chandra\ snapshot observations for a sample of 7 galaxies 
selected from the Arecibo OHM survey at \hbox{$z\approx0.13-0.22$} and with 
FIR luminosities in excess of 10$^{11}$~\lsun.
The principal results of this work are 
\pn$\bullet$
All of the observed sources are \xray\ weak; only one OHM, 
IRAS~FSC~03521$+$0028 ($z=0.15$) , was detected by \chandra, with 5 counts in 
the \hbox{0.5--8~keV} band. 
\pn$\bullet$
The \xray\ emission is consistent with that from star formation 
(as also suggested by the comparison with most of the literature OHMs 
with \xray\ constraints) and LLAGN; 
however, if an AGN is present, its contribution to the broad-band emission of 
OHM galaxies is probably modest. 
\pn$\bullet$
Under reasonable assumptions about the intrinsic \xray\ spectral shape, 
the observed count distribution from stacking analysis suggests absorption of 
$\approx10^{22}$~cm$^{-2}$.

\section*{Acknowledgments}
We acknowledge financial support from MIUR (COFIN grant \hbox{03-02-23}; 
CV and AC), \chandra\ \xray\ Center grant \hbox{G04-5104X} (CV and WNB), 
and NASA LTSA grant NAG5-13035 (WNB).   
The authors thank D. Alexander for providing us with the data 
shown in Fig.~3, S. Bianchi and G. Matt for sharing information 
on the OHM observed by \xmm\ before publication, C. Gruppioni for useful 
discussions, A. Tarchi for a 
careful reading of the manuscript, and the referee for his/her useful comments.

\end{document}